\documentclass[article,reprint,amsmath,amssymb,superscriptaddress,longbibliography]{revtex4-1}
\usepackage{graphicx}
\usepackage{dcolumn}
\usepackage{bm}
\usepackage[colorlinks,
            linkcolor=blue,
            anchorcolor=blue,
            citecolor=blue,
            ]{hyperref}
\usepackage{breakurl}
\usepackage{multirow}
\usepackage{enumitem}
\usepackage{color}
\usepackage[pagewise]{lineno}
\usepackage{tablefootnote}
\usepackage{threeparttable}
\usepackage{tabularx}
\usepackage{textcomp}
\setcitestyle{super} 

\begin{document}

\title {Thermally driven multiple sliding states and avalanche dynamics in rhombohedral Pb(TaSe$_2$)$_2$ nanoflakes}

\author{Zhuokai Xu}
\thanks{Equal contributions}
 \affiliation{Key Laboratory for Quantum Materials of Zhejiang Province, Department of Physics, School of Science and Research Center for Industries of the Future, Westlake University, Hangzhou 310030, P. R. China}
 \affiliation{Institute of Natural Sciences, Westlake Institute for Advanced Study, Hangzhou 310024, P. R. China}
 
\author{Xiaohui Yang}
\thanks{Equal contributions; Corresponding author}
\email{xiaohuiyang@cjlu.edu.cn}
 \affiliation{Department of Physics, China Jiliang University, Hangzhou 310018, Zhejiang, P. R. China}
 
 
 \affiliation{Key Laboratory for Quantum Materials of Zhejiang Province, Department of Physics, School of Science and Research Center for Industries of the Future, Westlake University, Hangzhou 310030, P. R. China}
 
\author{Zhefeng Lou}
 \affiliation{Key Laboratory for Quantum Materials of Zhejiang Province, Department of Physics, School of Science and Research Center for Industries of the Future, Westlake University, Hangzhou 310030, P. R. China}
 \affiliation{Institute of Natural Sciences, Westlake Institute for Advanced Study, Hangzhou 310024, P. R. China}

\author{Tian Le}
\affiliation{Center for Quantum Matter, School of Physics, Zhejiang University, Hangzhou 310058, China}

\author{Jialu Wang}

\affiliation{Hangzhou Key Laboratory of Quantum Matters, School of Physics, Hangzhou Normal University, Hangzhou 311121, China}

\author{Jiajun Ma}
 \affiliation{Key Laboratory for Quantum Materials of Zhejiang Province, Department of Physics, School of Science and Research Center for Industries of the Future, Westlake University, Hangzhou 310030, P. R. China}
 \affiliation{Institute of Natural Sciences, Westlake Institute for Advanced Study, Hangzhou 310024, P. R. China}

 \author{Ruiya Zhan}
 \affiliation{Key Laboratory for Quantum Materials of Zhejiang Province, Department of Physics, School of Science and Research Center for Industries of the Future, Westlake University, Hangzhou 310030, P. R. China}
 \affiliation{Institute of Natural Sciences, Westlake Institute for Advanced Study, Hangzhou 310024, P. R. China}

\author{Xiao Lin}
\thanks{Corresponding author}
 \email{linxiao@westlake.edu.cn}
 \affiliation{Key Laboratory for Quantum Materials of Zhejiang Province, Department of Physics, School of Science and Research Center for Industries of the Future, Westlake University, Hangzhou 310030, P. R. China}
 \affiliation{Institute of Natural Sciences, Westlake Institute for Advanced Study, Hangzhou 310024, P. R. China}

\date{\today}

\begin{abstract}

\noindent
Interlayer sliding in van der Waals materials offers a promising route to control functional properties, yet a comprehensive understanding of the complex collective domain dynamics governing these transitions remains challenging. 
Here, we investigate thermally driven sliding dynamics in Pb(TaSe$_2$)$_2$ nanoflakes, incorporating transport and atomic-resolution scanning transmission electron microscopy (STEM) measurements. Reducing the flake thickness induces multiple stepwise resistance jumps with enhanced hysteresis,  corresponding to transitions between stacking domains accompanied by ripplocations. Statistical analysis of the jumps reveals collective avalanche behavior of domain dynamics with universal power-law scaling, quantifying the domain motion. Harnessing these features, we demonstrate controllable, non-volatile multi-level resistive switching, including write/erase operations, via thermal cycling. These findings provide valuable insights into the sliding dynamics for various 2D materials, with significant implications for the development of slidetronics devices.
\end{abstract}

\maketitle
\noindent 
Interlayer sliding in van der Waals (vdW) materials provides a versatile platform for tailoring electronic and structural properties\cite{fox2023stacking,vizner2024sliding}, offering a pathway to modulate emergent intriguing phenomena, including  ferroelectricity\cite{fei2018ferroelectric,jindal2023coupled,wang2022interfacial,meng2022sliding,sui2023sliding,yasuda2021stacking}, superconductivity \cite{jindal2023coupled,chaudhary2024superconductivity}, magnetism \cite{li2019pressure,liu2020magnetoelectric} and nontrivial topological phases \cite{xiao2020berry,ren2022quantum,nataf2020domain,qian2023switchable}.
Sliding transitions can be driven by external stimuli such as electric fields\cite{meng2022sliding}, mechanical strain \cite{oviedo2015situ}, chemical intercalation\cite{muscher2021highly}, optical pumping\cite{sie2019ultrafast}, and temperature\cite{hart2023emergent,cha2023order}, enabling dynamic control over stacking configurations in systems like MoS$_2$\cite{oviedo2015situ,meng2022sliding},  WTe$_2$\cite{muscher2021highly,sie2019ultrafast} and MoTe$_2$\cite{hart2023emergent}. 
Despite significant progress, a fundamental understanding of the sliding mechanisms, particularly the complex dynamics of the resulting domain structures, remains elusive and merits further investigation\cite{liu2024sliding, fox2023stacking, sui2024atomic}.

Recently, theoretical investigations employing molecular dynamics simulations have explored the dynamic properties of domain walls in sliding ferroelectrics\cite{he2024ultrafast}. Experimentally, multiple stepwise resistance jumps have been observed in few-layered 3R-MoS$_2$\cite{meng2022sliding,chen2026deterministic} and other materials\cite{lv2024multiresistance, wang2022interfacial}, which was considered as a consequence of domain wall motion. In parallel, the domain walls have been directly visualized in other layered materials, such as few-layered MoTe$_2$\cite{hart2023emergent} and $\gamma$-InSe\cite{sui2024atomic}, via atomic-resolution scanning transmission electron microscopy (STEM). However, a comprehensive picture connecting the atomistic structure, dynamic evolution, and macroscopic transport signatures of these domains during sliding transitions is still missing.

\begin{figure*}[hbtp]
 \includegraphics[width=16cm]{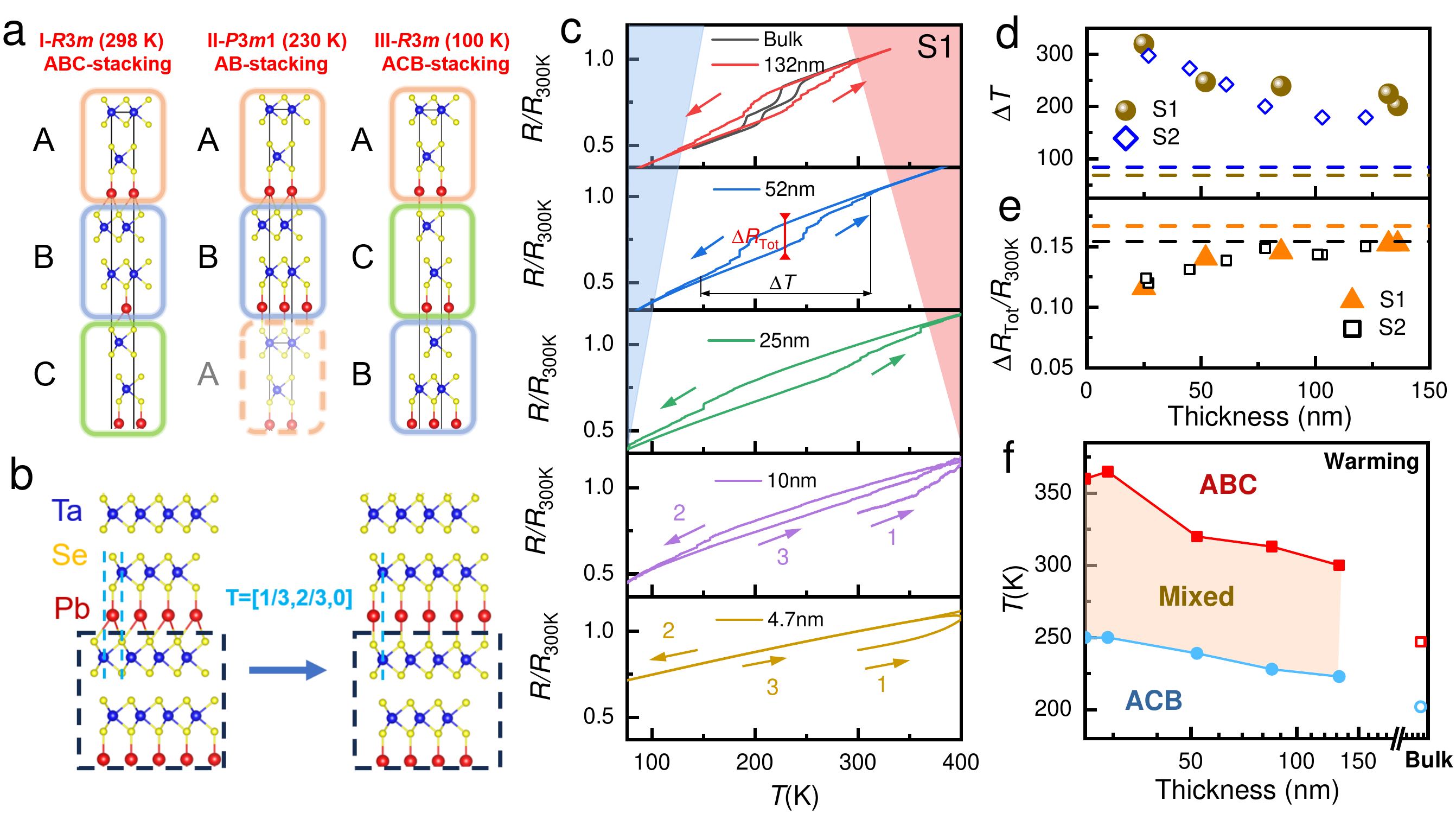}
  \setlength{\abovecaptionskip}{0pt}
 \caption{\textbf{Structural evolution and thickness-modulated electronic transport in Pb(TaSe$_2$)$_2$}  
 \textbf{a} Crystal structure evolution with three distinct phases in bulk Pb(TaSe$_2$)$_2$ : high-temperature 3R  (ABC stacking), intermediate Bernal (ABAB stacking), and low-temperature 3R (ACB stacking) phases. 
 \textbf{b} Interlayer sliding characterized by $\mathbf{T} = [1/3, 2/3, 0]$, accompanied by  local Pb-Se coordination reconfiguration from tetrahedral to dumbbell-like configuration. 
 \textbf{c} Temperature-dependence of normalized resistance ($R/R_{\text{300K}}$) for bulk S1 (topmost panel) and exfoliated nanoflakes of indicated thicknesses, obtained during thermal cycling. The arrows denote the temperature sweep direction. The numbers on 10 nm and 4.7 nm curves mark cycle segments. 
 \textbf{d-e} Thermal hysteresis width $\Delta T$ and normalized total stepwise resistance change $\Delta R_{\text{Tot}} / R_{300\text{K}}$ versus nanoflake thickness for S1 and S2 . Dashed lines denote the value from bulk. 
 \textbf{f} Temperature-thickness phase diagram for sample S1 obtain upon warming, derived from resistance jump regions in (c). The shaded region indicates the `` mixed '' phase  characterized by the step-like resistance behavior.
 }
  \label{Fig1}
\end{figure*}

To bridge this gap, we explore the two-stage intercalated 3R-TaSe$_2$ compound, Pb(TaSe$_2$)$_2$. The bulk single crystals exhibit unique, thermally induced two-step sliding transitions:  the building block (two TaSe$_2$ layers combined with a Pb layer) transforms from a high-temperature ABC stacking to an intermediate ABAB stacking, and finally to a low-temperature ACB stacking phase via a commensurate translation vector $\mathbf{T} = [1/3, 2/3, 0]$ \cite{yang2022commensurate}(Fig.~\ref{Fig1}a). This interlayer sliding induces a reconfiguration of the Pb–Se coordination, corresponding to a structural transition from tetrahedral to dumbbell‐like geometry, as depicted in Fig.~\ref{Fig1}b.

Here, we unravel the complex dynamics of thermally driven inter-layer sliding transitions in Pb(TaSe$_2$)$_2$ nanoflakes of reduced thickness, combining the measurements of transport and atomic-resolution STEM. Intriguingly, reducing the sample thickness drastically modifies the two-step resistance transition characteristic of bulk single crystals, leading to pronounced thermal hysteresis and multiple stepwise resistance jumps.
STEM directly visualizes the microscopic origin of the corresponding `` Mixed '' phase behavior, pointing to the spatial coexistence of the initial (ABC) and intermediate (ABAB) stacking domains, alongside structural features like ripplocations. Crucially, statistical analysis of the resistance jumps reveals the feature of ``collective avalanches'' in the domain reorganization that showcases universal, scale-invariant dynamics obeying a power-law scaling: $N(V)\sim V^{-(\alpha-1)}$ (with an exponent $\alpha \approx 1.8-2$) spanning over four orders of magnitude in estimated domain volume $V$.
Furthermore, leveraging the hysteretic, multi-step nature of the transition, we demonstrate robust, thermally controlled ``multi-level'' resistive switching, including ``write/erase'' operations, in a single device. Our findings not only advance the fundamental understanding of non-equilibrium dynamics in layered materials but also showcase pathways for harnessing these thermally controlled avalanche and switching phenomena in novel slidetronics and memory applications.

\vspace{0.5em}

\noindent \textbf{Results}

\noindent \textbf{Thickness dependence of resistivity} 

\noindent
Reducing the thickness ($d$) of Pb(TaSe$_2$)$_2$ single crystals reveals a dramatic change in their transport properties when compared to those of the bulk material. The nanoflakes are mechanically exfoliated from two bulk single crystals (S1 and S2), with the thickness down to 4.7 nm, approaching the unit cell of the ABC stacked phase.
Figure~\ref{Fig1}c shows the temperature-dependence of normalized resistance ($R/R_{\text{300K}}$) during the cooling and subsequent warming processes for S1 as well as the exfoliated nanoflakes of variable thicknesses.
Similar data for S2 are presented in Supplementary Fig.~1. Strikingly, the nanoflakes exhibit a series of sharp, stepwise resistance jumps signifying the structural transition, in stark contrast with the well-defined two-step transition observed in bulk.
This unique behavior is accompanied by a substantial enhancement of thermal hysteresis as marked by regions between two shaded areas in Fig.~\ref{Fig1}c, which remains robust across varying temperature ramp rates, as shown in Supplementary Fig.~2.
In the ultrathin limit (e.g., 10 nm and 4.7 nm flakes), the hysteresis becomes exceedingly pronounced that the cooling and warming curves exhibit no overlap within our measurement range (2 K-400 K), indicative of incomplete phase transitions.

\begin{figure*}[!thb]
 \includegraphics[width=16cm]{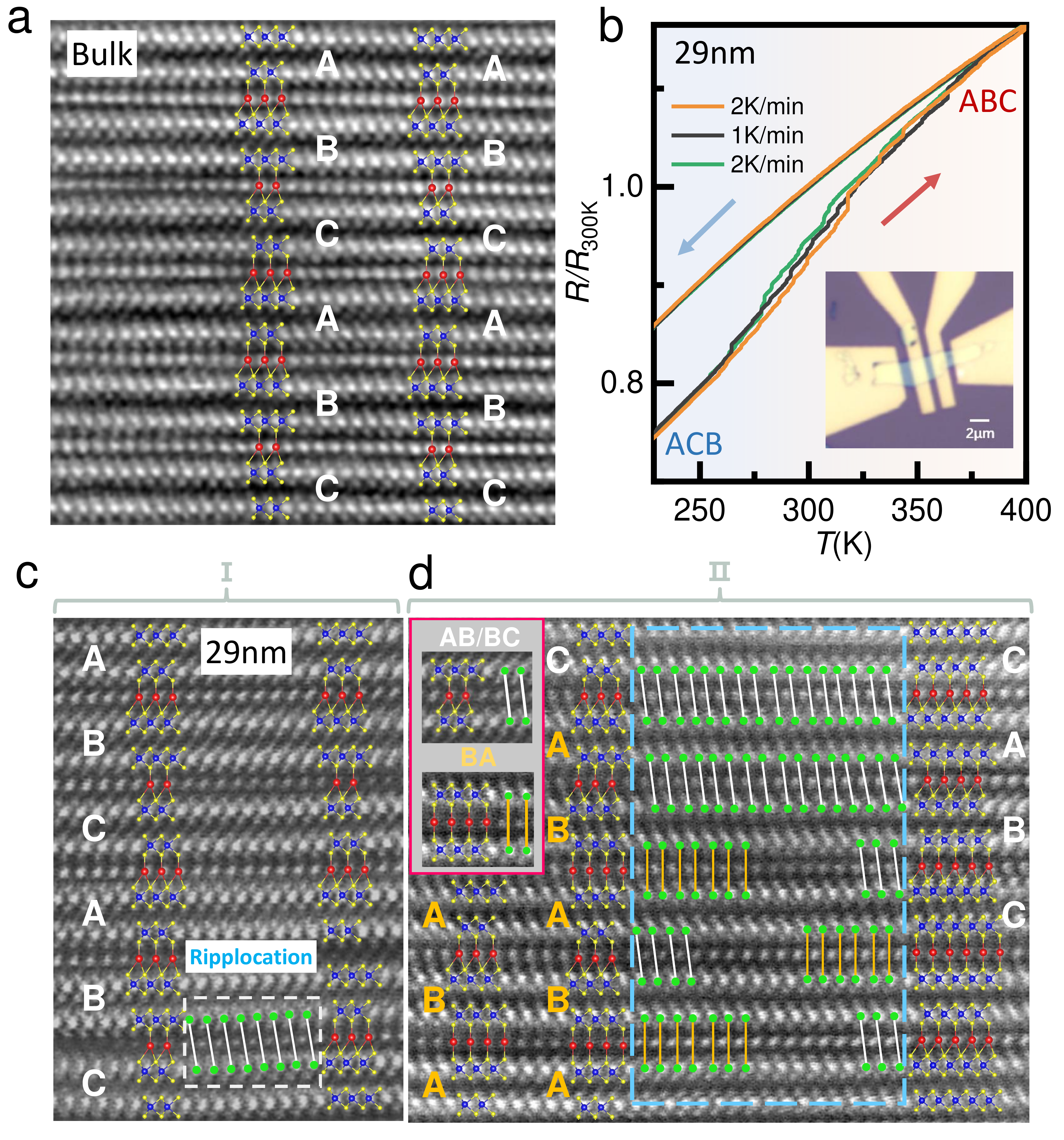}
  \setlength{\abovecaptionskip}{0pt}
  \caption{\textbf{Atomic-scale structure of the `` mixed '' phase.} 
  \textbf{a} Atomic-resolution STEM image of bulk Pb(TaSe$_2$)$_2$ at 300 K, showing the homogeneous ABC stacking configuration. 
  \textbf{b} Temperature dependence of normalized resistance ($R/R_{\text{300K}}$) for 29 nm-nanoflake, obtained during thermal cycling. The arrows denote the sweep direction. The inset is the optical image of device.
  \textbf{c-d} STEM image of  29 nm-nanoflake at 300 K. Panel (\textbf{c}) (Region I) displays an area with a homogeneous ABC stacking sequence and highlighted ripplocation (white dashed box). Panel (\textbf{d}) (Region II) shows the coexistence of ABC and ABAB stacking domains. The inset highlights the distinct AB/BC and BA type local stacking arrangements, as marked by colored guidelines (white and yellow).
  }
  
  \label{Fig2}
\end{figure*}

In Fig.~\ref{Fig1}d, the width of thermal hysteresis ($\Delta T$) increases as the thickness reduces, approaching a value of 300~K and beyond in thin flakes, which is among the widest hysteresis reported in crystalline materials\cite{lv2022unconventional}. Across the transition, the normalized total stepwise resistance change at 300~K ($\Delta R_{\text{Tot}} / R_{300\text{K}}$) is nearly constant in flakes exfoliated from two bulk samples (S1 and S2) at $d>75$~nm, decreasing slightly at smaller thicknesses, as shown in Fig.~\ref{Fig1}e. The temperature–thickness phase diagram for sample S1 during warming is shown in Fig.~\ref{Fig1}f. The colored region denotes the temperature range of resistance jumps, tentatively termed a ``mixed '' phase. 
Referring to the unique two-step transitions (ABC$\rightarrow$ABAB$\rightarrow$ACB) in bulk samples~\cite{yang2022commensurate}, the broadened ``mixed '' phase with multiple-step transitions in nanoflakes potentially suggests a spatially inhomogeneous transition pathway. This highlights the need for direct microscopic investigation of the ``mixed '' phase to clarify the structural evolution underlying the observed transport behavior.

\vspace{0.5em}
\noindent \textbf{Atomic-scale imaging of sliding phase transitions}

\begin{figure*}[!thb]
 \includegraphics[width=16cm]{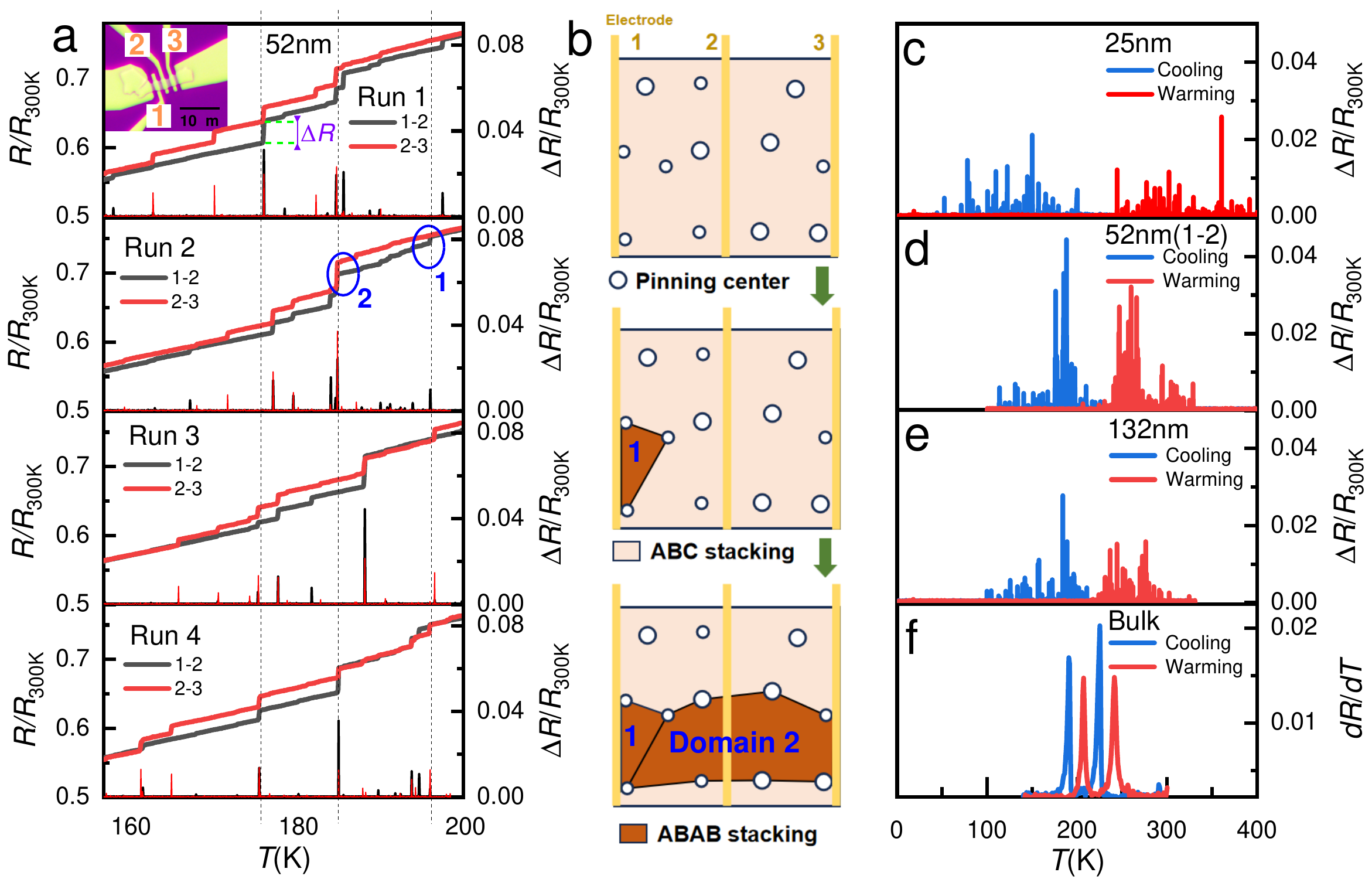}
  \setlength{\abovecaptionskip}{0pt}
  \caption{\textbf{Spatially inhomogeneous domain dynamics and statistical nature of resistance jumps in Pb(TaSe$_2$)$_2$ nanoflakes.} 
  \textbf{a} Temperature dependence of normalized resistance ($R/R_{\text{300K}}$) obtained between electrode pairs 1-2 (black curves) and 2-3 (red curves) on 52 nm-nanoflake during four consecutive cooling processes. Inset: optical image of the device. Vertical dashed lines mark reproducible resistance jump temperatures across different runs. Bar plots under each curve shows the magnitude ($\Delta R/R_{\text{300K}}$) of individual steps.
  \textbf{b} Schematics illustration of a proposed sequence of domain formation during cooling, corresponding to the events in Run 2. Top panel: Initial state with indicated pinning centers (circles). Middle panel: Formation of Domain 1. Bottom panel: Formation of Domain 2 extending across two regions. Yellow vertical lines represent electrodes; The areas of different colors represent different stacking phases (e.g., ABC, ABAB). 
  \textbf{c-f} Comparison of transition characteristics across different sample thicknesses. For bulk crystal, the plots show the temperature derivative of resistance ($dR/dT$) during cooling (blue) and warming (red). For nanoflakes of different thicknesses, the plots display the magnitude of normalized individual resistance steps ($\Delta R/R_{\text{300K}}$) during the cooling (blue) and warming (red) processes.
}
  \label{Fig3}
\end{figure*}

\noindent
To unravel the microscopic structure of the `` mixed '' phase, we performed atomic-resolution STEM on the cross-sectional specimens of Pb(TaSe$_2$)$_2$ prepared by focused ion beam (FIB) milling. Figure~\ref{Fig2}a presents the high-angle annular dark-field (HAADF)-STEM image of a bulk specimen at 300~K, which displays a homogeneous ABC stacking configuration, in good agreement with previous reports\cite{yang2022commensurate}. 

While for a 29-nm nanoflake, the resistance profile indicates that it resides in the `` mixed '' phase at 300 K as seen in Fig.~\ref{Fig2}b. 
The HAADF-STEM images of this specimen reveal a complex, spatially inhomogeneous structure consisting of two distinct regimes in Fig.~\ref{Fig2}c and 2d. Regime I is dominated by the original high-temperature ABC stacking, though some regions exhibit ripplocations (localized lattice distortions manifested as wavy interlayer displacements).  Ripplocations, often attributed to exfoliation-induced strain or interlayer sliding \cite{liang2023shear,wu2021sliding,kushima2015ripplocations,oviedo2015situ}, are commonly observed in vdW materials undergoing structural phase transitions \cite{sui2024atomic,wu2021sliding}.

Intriguingly, Regime II displays intricate configurations with the coexistence of high-temperature ABC and intermediate ABAB stacking. As illustrated in Fig.~\ref{Fig1}a, the AB and BC stacking segments in the ABC sequence exhibit laterally offset Ta atoms. This offset is highlighted by the inclined white lines, as exemplified in the solid rectangle of Fig.~\ref{Fig2}d. While the BA stacking segment, an exclusive feature of the ABAB sequence, exhibits vertically aligned Ta atoms, as indicated by the vertical yellow lines. This distinction enables clear resolution of ABAB-dominated domains on the left and ABC domains on the right in Fig.~\ref{Fig2}d. The two domain types are separated by an extended region of stacking disorder,  as  indicated by the dashed rectangle. The direct visualization of coexisting ABC and ABAB domains in the `` mixed '' phase of the 29 nm flake demonstrates the ABC$\rightarrow$ABAB phase transition, as observed in the bulk material\cite{yang2022commensurate}. The spatial evolution of microscopically coexisting nanoscale ABC and ABAB domains may give rise to the broadened multi-step resistive transition in thin flakes, as shown in Fig.~\ref{Fig1}c. This complex evolution is likely tied to the formation and pinning dynamics of ABAB domains within the ABC matrix.

\begin{figure*}[!thb]
 \includegraphics[width=16cm]{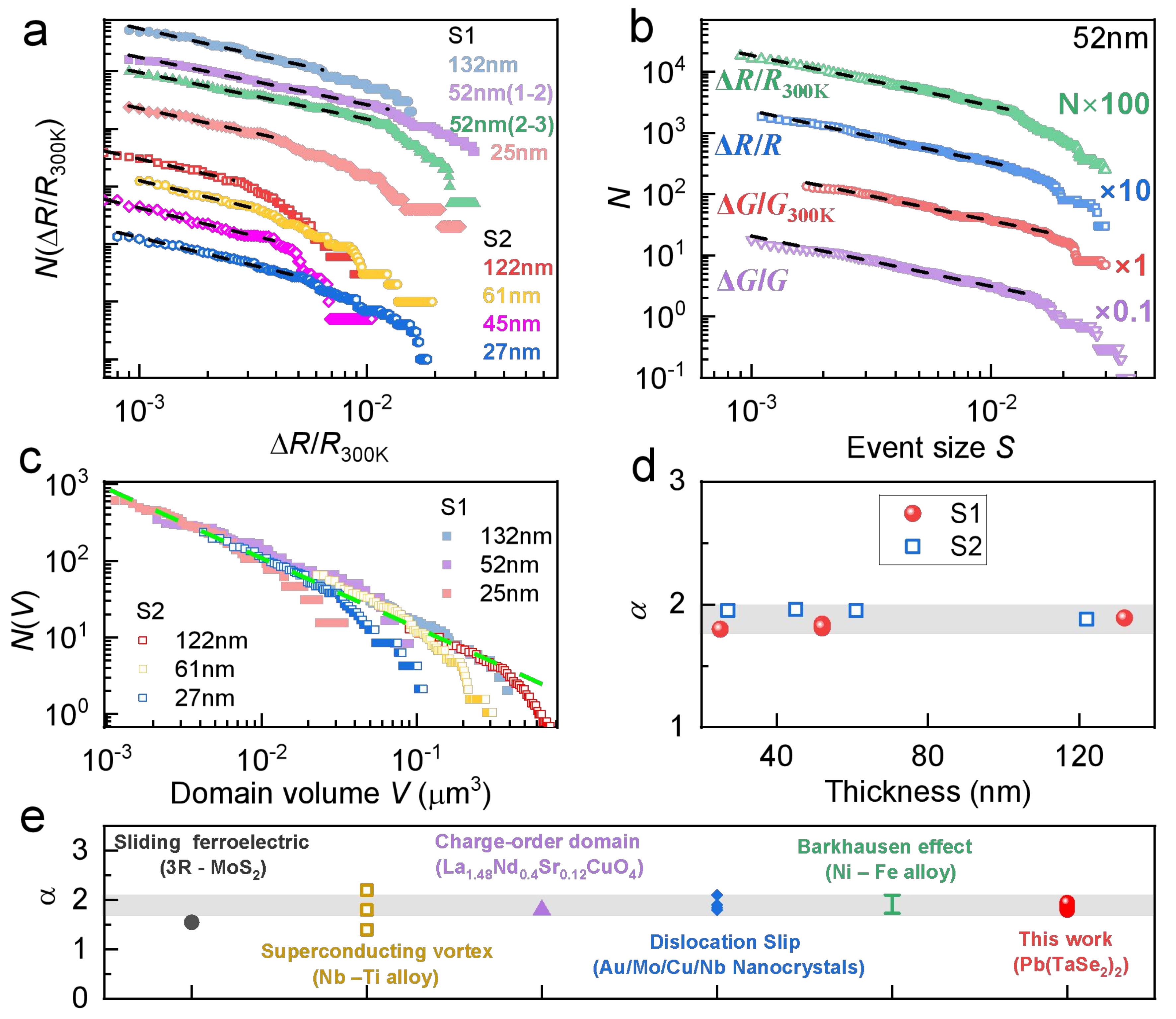}
  \caption{\textbf{Universal power-law scaling of sliding avalanches in Pb(TaSe$_2$)$_2$ nanoflakes.} 
  \textbf{a} Log-log plots of the cumulative number of resistance jumps $N(\Delta R/R_{\text{300K}})$ versus jump magnitude $\Delta R/R_{\text{300K}}$ for nanoflakes of various thickness exfoliated from S1 and S2. The curves are offset for clarity. The dashed black lines are power-law fits. 
  \textbf{b} Comparison of cumulative distributions, $N$, for  52 nm-nanoflake (S1) using four definitions of ``event size": normalization to the resistance/conductance at 300 K ($\Delta R/R_{\text{300K}}$, $\Delta G/G_{\text{300K}}$) and the local values just before each jump ($\Delta R/R$, $\Delta G/G$). Curves are shifted vertically by multiplicative factors for clarity. Similar slopes of the power-law fits (dashed black lines) confirm the robustness of the observed scaling behavior.
  \textbf{c} Log-log plot of the normalized cumulative avalanche number $N(V)$ versus estimated domain volume $V$ for S1 and S2 of various thicknesses, showing collapsed curves. 
  The green dashed line is a guide to the eye, highlighting the power-law scaling $N(V) \propto V^{-(\alpha-1)}$.
  \textbf{d} Extracted power-law exponent $\alpha$ as a function of nanoflake thickness for  S1 (red) and S2 (blue). The shaded region marks the range $\alpha \approx 1.8-2$. 
 \textbf{e} Comparison of $\alpha$ in this work with values reported for avalanche dynamics across diverse physical systems\cite{perkovic1995avalanches,friedman2012statistics,baity2018collective,liang2023shear,field1995superconducting}.
}
  \label{Fig4}
\end{figure*}

\vspace{0.5em}
\noindent \textbf{Sliding Dynamics in Pb(TaSe$_2$)$_2$}

\noindent
To gain a comprehensive understanding of the sliding phase transition, systematic temperature-dependent transport measurements were performed on different electrode pairs within a single nanoflake. Figure~\ref{Fig3}a presents the normalized resistance ($R/R_{\text{300K}}$) at terminals 1-2 and 2-3 ($V_\mathrm{12}$ and $V_\mathrm{23}$) for a 52-nm-thick flake, measured over four consecutive cooling cycles (Runs 1-4). An optical image of the device is shown in the inset. The resistance curves exhibit discrete resistance jumps, featuring the distinct stepwise sliding transitions. The magnitudes of these resistance steps (\(\Delta R/R_{300\text{K}}\)) are statistically summarized in the histograms positioned beneath each curve.

As shown in Fig.~\ref{Fig3}a,  $R/R_{\text{300K}}$ curves from the two terminals pairs are not identical within the the transition region ($\sim$150-200 K during cooling process), but collapses outside. For instance, as indicated by circle 1 in Run 2, a resistance jump occurs at $V_\mathrm{12}$, but not at $V_\mathrm{23}$. This suggests local ABAB domain organization in-between $V_\mathrm{12}$ during cooling, as  schematically illustrated in the middle panel of  Fig.~\ref{Fig3}b.  
While, as marked by circle 2, the resistance jump occurs simultaneously at $V_\mathrm{12}$ and $V_\mathrm{23}$, implying the formation or propagation of an extended domain spanning the two terminal regions, as illustrated in the bottom panel of Fig.~\ref{Fig3}b. This distinct behavior is highly linked to the random distribution of pinning centers within the exfoliated nanoflakes, that impedes domain wall motion~\cite{liang2023shear}. Additionally, the pinning strength varies spatially, therefore the domain wall in different regions may overcome the pining potential barrier for sliding at different temperatures~\cite{ko2023operando}. 
On the other hand, the repeated occurrence of resistance steps at similar temperature thresholds across multiple cooling processes (denoted by dashed lines in Fig.~\ref{Fig3}a) indicates that specific pinning centers consistently trap and release domain walls at reproducible thermal conditions.

In Fig.~\ref{Fig3}c-e, the  histograms of the resistance steps ($\Delta R/R_{\text{300K}}$) during cooling (blue) and  warming (red) are presented for nanoflakes with thicknesses of 25~nm, 52~nm, 132~nm, respectively. For comparison, Fig.~\ref{Fig3}f shows the temperature derivative of resistance ($dR/dT$) for the bulk crystal (S1), where the two sharp peaks signify well-defined two-step transitions. Notably,  as the flake thickness decreases, the distinct transitions in bulk evolve into multi-step processes distributed over an extended temperature window. Concurrently, thinner flakes exhibit more pronounced hysteresis during thermal cycling. Similar data from sample S2 is presented in Supplementary Fig.~1.
The observed enhancement of thermal hysteresis may stem from several factors,  including amplified surface/edge effects and substrate interactions (due to the increased surface-to-volume ratio)\cite{hart2023emergent,ko2023operando} as well as stronger pinning from intrinsic or exfoliation-induced defects in confined geometries.\cite{liang2023shear}.

\vspace{0.5em}
\noindent \textbf{Universal power law of sliding avalanches}

\noindent
The discrete resistance jumps observed during sliding transitions are indicative of avalanche dynamics, a phenomenon characteristic of systems responding to changing external conditions through discrete events spanning a broad range of sizes \cite{sethna2001crackling, papanikolaou2011universality}. 
A fundamental signature of such avalanche dynamics is the emergence of power-law scaling in the distribution of event sizes, typically expressed in the cumulative form $N(S) \propto S^{-(\alpha-1)}$\cite{oddershede1993self,liang2023shear,sethna2022power}, where $N(S)$ represents the number of events with a magnitude greater than ``event size'' $(S)$ and $\alpha$ is the scaling exponent (see Supplementary Note 1 for the statistical derivation) .
As shown in Fig.~\ref{Fig4}a, the cumulative distributions $N(\Delta R/R_{\text{300K}})$ for samples of various thicknesses exhibit a robust linear trend on a log-log scale, confirming the power-law behavior.
Notably, this scaling persists regardless of variations in temperature ramp rates and thermal cycling direction (Supplementary Fig.~3). Moreover, this scaling behavior is practically independent of how the ``event size'' is defined.
As illustrated for a representative 52~nm nanoflake in Fig.~\ref{Fig4}b and other thicknesses in Supplementary Fig.~4, the cumulative distributions derived from four distinct definitions ($S = \Delta R/R_{\text{300K}}$, $\Delta R/R$, $\Delta G/G_{\text{300K}}$, and $\Delta G/G$) consistently yield parallel power-law fits with nearly identical exponents. This consistency underscores an intrinsic dynamic feature of the system.

To relate these transport signatures to mesoscopic dynamics, we estimate an effective ``domain volume'' $V$ by assuming that each resistance jump $\Delta R$ is proportional to the volume of the region undergoing the stacking transition (see Supplementary Note 2 for details).
Remarkably, the data across all nanoflake thicknesses collapse onto a single universal power-law curve, $N(V) \propto V^{-(\alpha-1)}$, spanning four orders of magnitude (Fig.~\ref{Fig4}c).
$N(V)$ exhibits a downward deviation from the power-law scaling as each curve reaches a sample-specific volume limit, a behavior primarily attributed to finite-size  \cite{uhl2015universal,friedman2012statistics}.

As shown in Fig.~\ref{Fig4}d, the extracted scaling exponents $\alpha$ consistently fall within the range of 1.8 to 2.0. Notably, these values in Pb(TaSe$_2$)$_2$ are highly consistent with those reported for a wide array of physical systems exhibiting avalanche dynamics, including Barkhausen noise in magnets\cite{perkovic1995avalanches}, dislocation slip in metallic crystals\cite{friedman2012statistics}, charge-order domain dynamics in correlated electron systems\cite{baity2018collective} and others\cite{liang2023shear,field1995superconducting}, seen in Fig.~\ref{Fig4}e.

\begin{figure*}[!thb]
 \includegraphics[width=16cm]{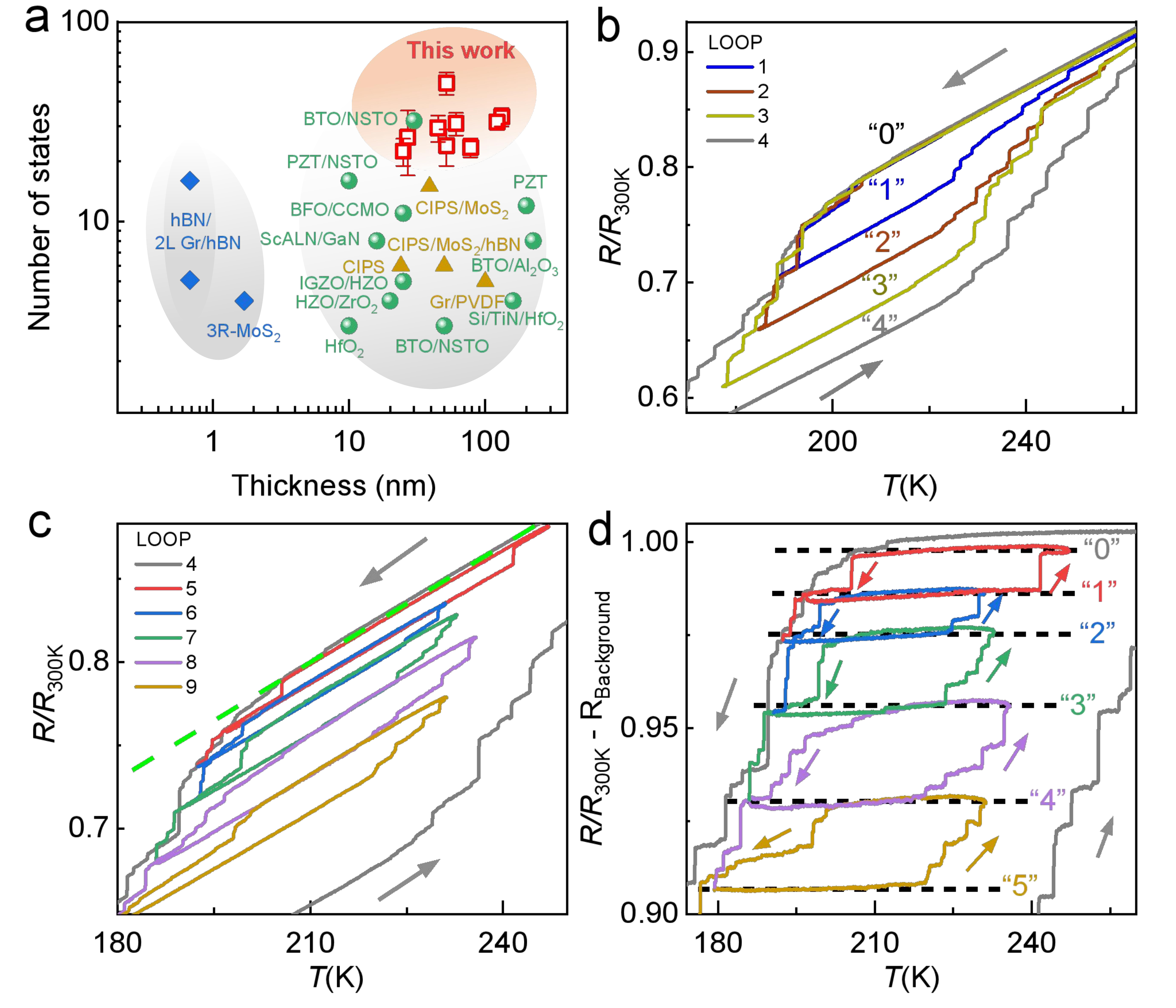}
  \setlength{\abovecaptionskip}{0pt}
  \caption{\textbf{Thermally modulated multi-level resistive states.} 
  \textbf{a} Numbers of physically distinct resistive states in this work, compared with various ferroelectric and sliding-based memory systems
  \cite{wang2026manipulating,lu2017multi, xu2019kinetic, luo2022high, zeng2021electric, soliman2023first, boyn2017learning, kim2020synergistic, ma2020sub, kumar2022high, geler2022ferroelectric, wang2023ultrathin, li2024realization, jandhyala2013graphene, li2023gate, wang2021van, wang2023moire, yan2023moire, bian2024developing}. Detailed information and corresponding references are listed in Supplementary Table 1.
  \textbf{b} Normalized resistance ($R/R_{\text{300K}}$) versus temperature, showing multiple distinct resistive states (labeled "0"–"4") achieved through four consecutive major thermal loops (Loops 1-4, starting from state "0") with different minimum temperatures during cooling. The arrows indicate temperature sweep direction. 
  \textbf{c} Normalized resistance ($R/R_{\text{300K}}$) versus temperature, showing switching between adjacent resistive states via minor thermal cycles within the main hysteresis loop.
 \textbf{d} Data from (c) replotted after subtraction of the a linear resistance background ($R_\text{Background}$), highlighting the  discrete resistance steps corresponding to programmable states (labeled "0"–"5"). Dashed lines are guides to the eye for the distinct resistance levels.  }
  \label{Fig5}
\end{figure*}

\vspace{0.5em}
\noindent \textbf{Multi-level operation for thermal memory}

\noindent
We first highlight the high-capacity storage potential of Pb(TaSe$_2$)$_2$ nanoflakes by benchmarking its accessible resistance levels against existing ferroelectric and sliding-based memory devices\cite{wang2026manipulating,lu2017multi, xu2019kinetic, luo2022high, zeng2021electric, soliman2023first, boyn2017learning, kim2020synergistic, ma2020sub, kumar2022high, geler2022ferroelectric, wang2023ultrathin, li2024realization, jandhyala2013graphene, li2023gate, wang2021van, wang2023moire, yan2023moire, bian2024developing}(Fig.\ref{Fig5}a). This remarkable state multiplicity far surpasses many traditional van der Waals materials, firmly positioning the system as a promising candidate for multi-level non-volatile memory. To explore its practical implementation, we performed controlled thermal cycling tests.

In each cycle, the nanoflake is cooled from the high-temperature phase, followed by warming initiated at specific, different temperatures ($T_\mathrm{w}$).  As shown in Fig.~\ref{Fig5}b, five distinct resistance states (marked as ``0''–``4'') are achieved by manually varying $T_\mathrm{w}$. The magnitude of each resistance state exhibits a monotonic dependence on $T_\mathrm{w}$, establishing a clear correlation between thermal history and resulting resistance configurations.
Fig~\ref{Fig5}c demonstrates a controllable switch between adjacent resistance states by thermal cycling within narrower temperature ranges.
To enhance visualization of discrete states, the data in Fig.~\ref{Fig5}c was subtracted by a $T$-linear background resistance (dashed lines). The resulting curves in Fig.~\ref{Fig5}d clearly resolve at least six switchable resistance plateaus (``0''–``5''), highlighting the reproducibility of the thermal-driven processes. Furthermore,  Supplementary Fig.~5 shows that the established resistance state is robust against substantial DC bias currents (up to 2 mA) and persist over a long duration ($10^5$~s), confirming their non-volatile nature.

The thermal-driven multi-level resistive switch opens up new possibilities for designing functional devices, leveraging thermal hysteresis and domain dynamics as controllable degrees of freedom.  The programmability  of discrete sliding resistance states via tailored thermal protocols allows for potential applications in non-volatile multi-level memory~\cite{xu2021non,wang2025composition,cai20258,you2014organic,rios2015integrated} and neuromorphic thermal sensors \cite{mallik2023thermally,wang2025composition}, wherein thermal manipulation could serve as an energy-efficient ``write/erase" operation, complementary to electrical addressing \cite{hyun2016multi,chen2022reproducible}. 

\vspace{0.5em}
\noindent \textbf{Discussion}

\noindent 
Our integrated study  uncovers the intricate interplay between dimensionality and domain dynamics in the thermally driven interlayer sliding processes within Pb(TaSe$_2$)$_2$ nanoflakes. The thickness reduction fundamentally alters the transition, leading to a `` mixed '' phase that underlies the multiple resistance jumps and extended thermal hysteresis. 
It enables functional thermal control of robust, non-volatile multi-level resistive switching for neuromorphic applications~\cite{xu2021non, mallik2023thermally}. 
To understand the microscopic origin of these dynamics, 
we argue that the observed phenomena are dominated by the unique pinning effect in reduced dimensions, where local defects  and strain fluctuations create spatially inhomogeneous energy barriers that hinder coherent interlayer sliding. These barriers stabilize metastable domain configurations and drive intermittent depinning processes\cite{liang2023shear, ko2023operando}.  
A subsequent quantitative analysis captures mesoscopic domain dynamics governed by collective avalanche events, characterized by a universal scaling behaviour. 
In essence, this work provides a quantitative insight into the collective domain dynamics governing sliding phase transitions, bridging the gap between microscopic structure (phase coexistence), mesoscopic dynamics (avalanches), and macroscopic transport.

\vspace{0.5em}
\noindent\textbf{Methods}

\noindent\textbf{Device fabrication}

\noindent 
Thin Pb(TaSe$_{2}$)$_{2}$ flakes were mechanically exfoliated from bulk single crystals and transferred onto silicon substrates capped with a 300-nm-thick SiO$_{2}$ layer. 
The samples were spin-coated with poly(methyl methacrylate) (PMMA) resist, and electrode patterns were subsequently defined using standard electron-beam lithography (Tescan Vega LMS). 
The patterns were developed in a methyl isobutyl ketone/isopropyl alcohol (MIBK:IPA = 1:3) solution for 60~s, followed by an IPA rinse. 
To ensure high-quality ohmic contacts, the flake surfaces were treated with Ar plasma cleaning prior to metal deposition. 
Finally, Ti/Au (5~nm/90~nm) contacts were deposited using electron-beam evaporation.

\vspace{0.5em}
\noindent\textbf{Thickness characterization and transport measurements}

\noindent
The flake thickness was determined using atomic force microscopy (Asylum Research Jupiter XR, Oxford Instruments). The transport measurements were performed on a physical property measurement system (Quantum Design) equipped with SR830 Lock-In Amplifier.

\vspace{0.5em}
\noindent\textbf{Scanning transmission electron microscopy}

\noindent
HAADF-STEM were carried out on a Double Cs-Corrected microscope (Spectra Ultra, Thermo Fisher Scientific) operated at 300 kV. TEM lamellas were prepared on a dual-beam Focused Ion Beam Scanning Electron Microscope system (Helios 5 UX, Thermo Fisher Scientific).

\vspace{0.5em}
\noindent\textbf{Data availability}

\noindent The data that support the findings of this study are included in this article and its supplementary information file. The datasets are available from the corresponding author upon reasonable request.

\vspace{0.5em}
\noindent \textbf{References}
\bibliographystyle{naturemag}

\bibliography{Pb124nano}

@article{fox2023stacking,
  title={Stacking order engineering of two-dimensional materials and device applications},
  author={Fox, Carter and Mao, Yulu and Zhang, Xiang and Wang, Ying and Xiao, Jun},
  journal={Chem. Rev.},
  volume={124},
  number={4},
  pages={1862--1898},
  year={2023},
  publisher={ACS Publications}
}

@article{vizner2024sliding,
  title={Sliding van der Waals polytypes},
  author={Vizner Stern, Maayan and Salleh Atri, Simon and Ben Shalom, Moshe},
  journal={Nat. Rev. Phys.},
  volume={7},
  number={1},
  pages={50--61},
  year={2025},
  publisher={Nature Publishing Group UK London}
}

@article{fei2018ferroelectric,
  title={Ferroelectric switching of a two-dimensional metal},
  author={Fei, Zaiyao and Zhao, Wenjin and Palomaki, Tauno A and Sun, Bosong and Miller, Moira K and Zhao, Zhiying and Yan, Jiaqiang and Xu, Xiaodong and Cobden, David H},
  journal={Nature},
  volume={560},
  number={7718},
  pages={336--339},
  year={2018},
  publisher={Nature Publishing Group UK London}
}

@article{jindal2023coupled,
  title={Coupled ferroelectricity and superconductivity in bilayer $\mathrm{T_{d}}$-$\mathrm{MoTe}_2$},
  author={Jindal, Apoorv and Saha, Amartyajyoti and Li, Zizhong and Taniguchi, Takashi and Watanabe, Kenji and Hone, James C and Birol, Turan and Fernandes, Rafael M and Dean, Cory R and Pasupathy, Abhay N and Rhodes, Daniel A},
  journal={Nature},
  volume={613},
  number={7942},
  pages={48--52},
  year={2023},
  publisher={Nature Publishing Group UK London}
}

@article{wang2022interfacial,
  title={Interfacial ferroelectricity in rhombohedral-stacked bilayer transition metal dichalcogenides},
  author={Wang, Xirui and Yasuda, Kenji and Zhang, Yang and Liu, Song and Watanabe, Kenji and Taniguchi, Takashi and Hone, James and Fu, Liang and Jarillo-Herrero, Pablo},
  journal={Nat. Nanotechnol.},
  volume={17},
  number={4},
  pages={367--371},
  year={2022},
  publisher={Nature Publishing Group UK London}
}

@article{meng2022sliding,
  title={Sliding induced multiple polarization states in two-dimensional ferroelectrics},
  author={Meng, Peng and Wu, Yaze and Bian, Renji and Pan, Er and Dong, Biao and Zhao, Xiaoxu and Chen, Jiangang and Wu, Lishu and Sun, Yuqi and Fu, Qundong 
and Liu, Qing
and Shi, Dong
and Zhang, Qi
and Zhang, Yong-Wei
and Liu, Zheng
and Liu, Fucai
},
  journal={Nat. Commun.},
  volume={13},
  number={1},
  pages={7696},
  year={2022},
  publisher={Nature Publishing Group UK London}
}

@article{sui2023sliding,
  title={Sliding ferroelectricity in van der Waals layered $\gamma$-$\mathrm{InSe}$ semiconductor},
  author={Sui, Fengrui and Jin, Min and Zhang, Yuanyuan and Qi, Ruijuan and Wu, Yu-Ning and Huang, Rong and Yue, Fangyu and Chu, Junhao},
  journal={Nat. Commun.},
  volume={14},
  number={1},
  pages={36},
  year={2023},
  publisher={Nature Publishing Group UK London}
}

@article{yasuda2021stacking,
  title={Stacking-engineered ferroelectricity in bilayer boron nitride},
  author={Yasuda, Kenji and Wang, Xirui and Watanabe, Kenji and Taniguchi, Takashi and Jarillo-Herrero, Pablo},
  journal={Science},
  volume={372},
  number={6549},
  pages={1458--1462},
  year={2021},
  publisher={American Association for the Advancement of Science}
}

@article{chaudhary2024superconductivity,
  title={Superconductivity from domain wall fluctuations in sliding ferroelectrics},
  author={Chaudhary, Gaurav and Martin, Ivar},
  journal={Phys. Rev. Lett.},
  volume={133},
  number={24},
  pages={246001},
  year={2024},
  publisher={APS}
}

@article{li2019pressure,
  title={Pressure-controlled interlayer magnetism in atomically thin $\mathrm{CrI}_3$},
  author={Li, Tingxin and Jiang, Shengwei and Sivadas, Nikhil and Wang, Zefang and Xu, Yang and Weber, Daniel and Goldberger, Joshua E and Watanabe, Kenji and Taniguchi, Takashi and Fennie, Craig J and Fai Mak, Kin and  Shan, Jie
},
  journal={Nat. Mater.},
  volume={18},
  number={12},
  pages={1303--1308},
  year={2019},
  publisher={Nature Publishing Group UK London}
}

@article{liu2020magnetoelectric,
  title={Magnetoelectric coupling in multiferroic bilayer VS 2},
  author={Liu, Xingen and Pyatakov, Alexander P and Ren, Wei},
  journal={Phys. Rev. Lett.},
  volume={125},
  number={24},
  pages={247601},
  year={2020},
  publisher={APS}
}

@article{xiao2020berry,
  title={Berry curvature memory through electrically driven stacking transitions},
  author={Xiao, Jun and Wang, Ying and Wang, Hua and Pemmaraju, CD and Wang, Siqi and Muscher, Philipp and Sie, Edbert J and Nyby, Clara M and Devereaux, Thomas P and Qian, Xiaofeng and Zhang, Xiang and Lindenberg, Aaron M.},
  journal={Nat. Phys.},
  volume={16},
  number={10},
  pages={1028--1034},
  year={2020},
  publisher={Nature Publishing Group UK London}
}

@article{ren2022quantum,
  title = {Quantum phase transitions driven by sliding in bilayer  $\mathrm{MnBi}_{2}\mathrm{Te}_{4}$},
  author = {Ren, Yi and Ke, Shasha and Lou, Wen-Kai and Chang, Kai},
  journal = {Phys. Rev. B},
  volume = {106},
  issue = {23},
  pages = {235302},
  numpages = {6},
  year = {2022},
  month = {Dec},
  publisher = {American Physical Society},
}

@article{nataf2020domain,
  title={Domain-wall engineering and topological defects in ferroelectric and ferroelastic materials},
  author={Nataf, GF and Guennou, Mael and Gregg, JM and Meier, D and Hlinka, J and Salje, EKH and Kreisel, Jens},
  journal={Nat. Rev. Phys.},
  volume={2},
  number={11},
  pages={634--648},
  year={2020},
  publisher={Nature Publishing Group UK London}
}

@article{qian2023switchable,
  title={Switchable band topology and geometric current in sliding bilayer elemental ferroelectric},
  author={Qian, Zhuang and Gong, Zhihao and Li, Jian and Wang, Hua and Liu, Shi},
  journal={Chin. Phys. Lett.},
  volume={42},
  number={7},
  pages={070715},
  year={2025},
  publisher={Chinese Physical Society and IOP Publishing Ltd}
}

@article{oviedo2015situ,
  title={In situ $\mathrm{TEM}$ characterization of shear-stress-induced interlayer sliding in the cross section view of molybdenum disulfide},
  author={Oviedo, Juan Pablo and KC, Santosh and Lu, Ning and Wang, Jinguo and Cho, Kyeongjae and Wallace, Robert M and Kim, Moon J},
  journal={ACS Nano},
  volume={9},
  number={2},
  pages={1543--1551},
  year={2015},
  publisher={ACS Publications}
}

@article{muscher2021highly,
  title={Highly efficient uniaxial in-plane stretching of a 2$\mathrm{D}$ material via ion insertion},
  author={Muscher, Philipp K. and Rehn, Daniel A. and Sood, Aditya and Lim, Kipil and Luo, Duan and Shen, Xiaozhe and Zajac, Marc and Lu, Feiyu and Mehta, Apurva and Li, Yiyang and Wang, Xijie and Reed, Evan J. and Chueh, William C. and Lindenberg, Aaron M.},
  journal={Adv. Mater.},
  volume={33},
  number={37},
  pages={2101875},
  year={2021},
  publisher={Wiley Online Library}
}

@article{sie2019ultrafast,
  title={An ultrafast symmetry switch in a Weyl semimetal},
  author={Sie, Edbert J and Nyby, Clara M and Pemmaraju, CD and Park, Su Ji and Shen, Xiaozhe and Yang, Jie and Hoffmann, Matthias C and Ofori-Okai, BK and Li, Renkai and Reid, Alexander H
and Weathersby, Stephen
and Mannebach, Ehren
and Finney, Nathan
and Rhodes, Daniel
and Chenet, Daniel
and Antony, Abhinandan
and Balicas, Luis
and Hone, James
and Devereaux, Thomas P.
and Heinz, Tony F.
and Wang, Xijie
and Lindenberg, Aaron M.
},
  journal={Nature},
  volume={565},
  number={7737},
  pages={61--66},
  year={2019},
  publisher={Nature Publishing Group UK London}
}

@article{hart2023emergent,
  title={Emergent layer stacking arrangements in c-axis confined $\mathrm{MoTe_{2}}$},
  author={Hart, James L.
and Bhatt, Lopa
and Zhu, Yanbing
and Han, Myung-Geun
and Bianco, Elisabeth
and Li, Shunran
and Hynek, David J.
and Schneeloch, John A.
and Tao, Yu
and Louca, Despina
and Guo, Peijun
and Zhu, Yimei
and Jornada, Felipe
and Reed, Evan J.
and Kourkoutis, Lena F.
and Cha, Judy J.
},
  journal={Nat. Commun.},
  volume={14},
  number={1},
  pages={4803},
  year={2023},
  publisher={Nature Publishing Group UK London}
}

@article{cha2023order,
  title={Order-disorder phase transition driven by interlayer sliding in lead iodides},
  author={Cha, Seyeong and Lee, Giyeok and Lee, Sol and Ryu, Sae Hee and Sohn, Yeongsup and An, Gijeong and Kang, Changmo and Kim, Minsu and Kim, Kwanpyo and Soon, Aloysius and Kim, Keun Su},
  journal={Nat. Commun.},
  volume={14},
  number={1},
  pages={1981},
  year={2023},
  publisher={Nature Publishing Group UK London}
}

@article{liu2024sliding,
  title={Sliding ferroelectricity: theory, experiment, and potential applications},
  author={Liu, YANG and Meng-Hao, WU},
  journal={PHYSICS},
  volume={53},
  number={11},
  pages={741--750},
  year={2024}
}

@article{sui2024atomic,
  title={Atomic-level polarization reversal in sliding ferroelectric semiconductors},
  author={Sui, Fengrui
and Li, Haoyang
and Qi, Ruijuan
and Jin, Min
and Lv, Zhiwei
and Wu, Menghao
and Liu, Xuechao
and Zheng, Yufan
and Liu, Beituo
and Ge, Rui
and Wu, Yu-Ning
and Huang, Rong
and Yue, Fangyu
and Chu, Junhao
and Duan, Chungang
},
  journal={Nat. Commun.},
  volume={15},
  number={1},
  pages={3799},
  year={2024},
  publisher={Nature Publishing Group UK London}
}

@article{he2024ultrafast,
  title={Ultrafast switching dynamics of the ferroelectric order in stacking-engineered ferroelectrics},
  author={He, Ri and Zhang, Bingwen and Wang, Hua and Li, Lei and Tang, Ping and Bauer, Gerrit and Zhong, Zhicheng},
  journal={Acta Mater.},
  volume={262},
  pages={119416},
  year={2024},
  publisher={Elsevier}
}

@article{chen2026deterministic,
  title={Deterministic 1$\mathrm{D}$ Domain Wall Motion with Nucleation-Free Nature in Sliding Ferroelectric Switching},
  author={Chen, Jiangang and Ke, Changming and Bian, Renji and Pan, Er and Kun, Peter and Li, Zefen and Dong, Biao and Yang, Fan and Liu, Qing and Tapaszto, Levente and Luo, Xiao and Liu, Shi and Liu, Fucai},
  journal={Phys. Rev. X},
  volume={16},
  number={1},
  pages={011066},
  year={2026},
  publisher={APS}
}

@article{lv2024multiresistance,
  title={Multiresistance states in ferro-and antiferroelectric trilayer boron nitride},
  author={Lv, Ming and Wang, Jiulong and Tian, Ming and Wan, Neng and Tong, Wenyi and Duan, Chungang and Xue, Jiamin},
  journal={Nat. Commun.},
  volume={15},
  number={1},
  pages={295},
  year={2024},
  publisher={Nature Publishing Group UK London}
}

@article{yang2022commensurate,
  title={Commensurate stacking phase transitions in an intercalated transition metal dichalcogenide},
  author={Yang, Xiaohui
and Bao, Jin-Ke
and Lou, Zhefeng
and Li, Peng
and Jiang, Chenxi
and Wang, Jialu
and Sun, Tulai
and Liu, Yabin
and Guo, Wei
and Ramakrishnan, Sitaram
and Kotla, Surya Rohith
and Tolkiehn, Martin
and Paulmann, Carsten
and Cao, Guang-Han
and Nie, Yuefeng
and Li, Wenbin
and Liu, Yang
and van Smaalen, Sander
and Lin, Xiao
and Xu, Zhu-An
},
  journal={Adv. Mater.},
  volume={34},
  number={6},
  pages={2108550},
  year={2022},
  publisher={Wiley Online Library}
}

@article{lv2022unconventional,
  title={Unconventional hysteretic transition in a charge density wave},
  author={Lv, B. Q. and Zong, Alfred and Wu, D. and Rozhkov, A. V. and Fine, Boris V. and Chen, Su-Di and Hashimoto, Makoto and Lu, Dong-Hui and Li, M. and Huang, Y.-B. and Ruff, Jacob P. C. and Walko, Donald A. and Chen, Z. H. and Hwang, Inhui and Su, Yifan and Shen, Xiaozhe and Wang, Xirui and Han, Fei and Po, Hoi Chun and Wang, Yao and Jarillo-Herrero, Pablo and Wang, Xijie and Zhou, Hua and Sun, Cheng-Jun and Wen, Haidan and Shen, Zhi-Xun and Wang, N. L. and Gedik, Nuh},
  journal={Phys. Rev. Lett.},
  volume={128},
  number={3},
  pages={036401},
  year={2022},
  publisher={APS}
}

@article{liang2023shear,
  title={Shear strain-induced two-dimensional slip avalanches in rhombohedral $\mathrm{MoS_{2}}$},
  author={Liang, Jing and Yang, Dongyang and Xiao, Yunhuan and Chen, Sean and Dadap, Jerry I and Rottler, Joerg and Ye, Ziliang},
  journal={Nano Lett.},
  volume={23},
  number={15},
  pages={7228--7235},
  year={2023},
  publisher={ACS Publications}
}

@article{wu2021sliding,
  title={Sliding ferroelectricity in 2$\mathrm{D}$ van der Waals materials: Related physics and future opportunities},
  author={Wu, Menghao and Li, Ju},
  journal={Proc. Natl. Acad. Sci.},
  volume={118},
  number={50},
  pages={e2115703118},
  year={2021},
  publisher={National Acad Sciences}
}

@article{kushima2015ripplocations,
  title={Ripplocations in van der Waals layers},
  author={Kushima, Akihiro and Qian, Xiaofeng and Zhao, Peng and Zhang, Sulin and Li, Ju},
  journal={Nano Lett.},
  volume={15},
  number={2},
  pages={1302--1308},
  year={2015},
  publisher={ACS Publications}
}

@article{perkovic1995avalanches,
  title={Avalanches, Barkhausen noise, and plain old criticality},
  author={Perkovi{\'c}, Olga and Dahmen, Karin and Sethna, James P},
  journal={Phys. Rev. Lett.},
  volume={75},
  number={24},
  pages={4528},
  year={1995},
  publisher={APS}
}

@article{friedman2012statistics,
  title={Statistics of dislocation slip avalanches in nanosized single crystals show tuned critical behavior predicted by a simple mean field model},
  author={Friedman, Nir and Jennings, Andrew T and Tsekenis, Georgios and Kim, Ju-Young and Tao, Molei and Uhl, Jonathan T and Greer, Julia R and Dahmen, Karin A},
  journal={Phys. Rev. Lett.},
  volume={109},
  number={9},
  pages={095507},
  year={2012},
  publisher={APS}
}

@article{baity2018collective,
  title={Collective dynamics and strong pinning near the onset of charge order in 
  $\mathrm{La_{1.48}Nd_{0.4}Sr_{0.12}CuO_{4}}$},
  author={Baity, PG and Sasagawa, T and Popovi{\'c}, Dragana},
  journal={Phys. Rev. Lett.},
  volume={120},
  number={15},
  pages={156602},
  year={2018},
  publisher={APS}
}

@article{field1995superconducting,
  title={Superconducting vortex avalanches},
  author={Field, Stuart and Witt, Jeff and Nori, Franco and Ling, Xinsheng},
  journal={Phys. Rev. Lett.},
  volume={74},
  number={7},
  pages={1206},
  year={1995},
  publisher={APS}
}

@article{ko2023operando,
  title={Operando electron microscopy investigation of polar domain dynamics in twisted van der Waals homobilayers},
  author={Ko, Kahyun
and Yuk, Ayoung
and Engelke, Rebecca
and Carr, Stephen
and Kim, Junhyung
and Park, Daesung
and Heo, Hoseok
and Kim, Hyun-Mi
and Kim, Seul-Gi
and Kim, Hyeongkeun
and Taniguchi, Takashi
and Watanabe, Kenji
and Park, Hongkun
and Kaxiras, Efthimios
and Yang, Sang Mo
and Kim, Philip
and Yoo, Hyobin
},
  journal={Nat. Mater.},
  volume={22},
  number={8},
  pages={992--998},
  year={2023},
  publisher={Nature Publishing Group UK London}
}

@article{sethna2001crackling,
  title={Crackling noise},
  author={Sethna, James P and Dahmen, Karin A and Myers, Christopher R},
  journal={Nature},
  volume={410},
  number={6825},
  pages={242--250},
  year={2001},
  publisher={Nature Publishing Group UK London}
}

@article{papanikolaou2011universality,
  title={Universality beyond power laws and the average avalanche shape},
  author={Papanikolaou, Stefanos and Bohn, Felipe and Sommer, Rubem Luis and Durin, Gianfranco and Zapperi, Stefano and Sethna, James P},
  journal={Nat. Phys.},
  volume={7},
  number={4},
  pages={316--320},
  year={2011},
  publisher={Nature Publishing Group UK London}
}

@article{oddershede1993self,
  title={Self-organized criticality in fragmenting},
  author={Oddershede, Lene and Dimon, Peter and Bohr, Jakob},
  journal={Phys. Rev. Lett.},
  volume={71},
  number={19},
  pages={3107},
  year={1993},
  publisher={APS}
}

@article{sethna2022power,
  title={Power laws in physics},
  author={Sethna, James P},
  journal={Nature Reviews Physics},
  volume={4},
  number={8},
  pages={501--503},
  year={2022},
  publisher={Nature Publishing Group UK London}
}

@article{uhl2015universal,
  title={Universal quake statistics: From compressed nanocrystals to earthquakes},
  author={Uhl, Jonathan T.
and Pathak, Shivesh
and Schorlemmer, Danijel
and Liu, Xin
and Swindeman, Ryan
and Brinkman, Braden A. W.
and LeBlanc, Michael
and Tsekenis, Georgios
and Friedman, Nir
and Behringer, Robert
and Denisov, Dmitry
and Schall, Peter
and Gu, Xiaojun
and Wright, Wendelin J.
and Hufnagel, Todd
and Jennings, Andrew
and Greer, Julia R.
and Liaw, P. K.
and Becker, Thorsten
and Dresen, Georg
and Dahmen, Karin A.
},
  journal={Sci. Rep.},
  volume={5},
  number={1},
  pages={16493},
  year={2015},
  publisher={Nature Publishing Group UK London}
}

@article{wang2026manipulating,
  title={Manipulating thousands of non-volatile polarization states within one sliding ferroelectric transistor at room temperature},
  author={Wang, Xiaofan and Chen, Xiaokai and Long, Yuyang and Liu, Jinguo and Lin, Fanrong and Yin, Jun and Liu, Yanpeng and Guo, Wanlin},
  journal={Nat. Electron.},
  pages={1--9},
  year={2026},
  publisher={Nature Publishing Group UK London}
}

@article{lu2017multi,
  title={Multi-nonvolatile state resistive switching arising from ferroelectricity and oxygen vacancy migration},
  author={L{\"u}, Weiming and Li, Changjian and Zheng, Limei and Xiao, Juanxiu and Lin, Weinan and Li, Qiang and Wang, Xiao Renshaw and Huang, Zhen and Zeng, Shengwei and Han, Kun and  Zhou, Wenxiong
 and  Zeng, Kaiyang
 and  Chen, Jingsheng
 and  Ariando
 and  Cao, Wenwu
 and  Venkatesan, Thirumalai
  },
  journal={Adv. Mater.},
  volume={29},
  number={24},
  pages={1606165},
  year={2017},
  publisher={Wiley Online Library}
}

@article{xu2019kinetic,
  title={Kinetic control of tunable multi-state switching in ferroelectric thin films},
  author={Xu, R.
and Liu, S.
and Saremi, S.
and Gao, R.
and Wang, J. J.
and Hong, Z.
and Lu, H.
and Ghosh, A.
and Pandya, S.
and Bonturim, E.
and Chen, Z. H.
and Chen, L. Q.
and Rappe, A. M.
and Martin, L. W.


  },
  journal={Nat. Commun.},
  volume={10},
  number={1},
  pages={1282},
  year={2019},
  publisher={Nature Publishing Group UK London}
}

@article{luo2022high,
  title={High-precision and linear weight updates by subnanosecond pulses in ferroelectric tunnel junction for neuro-inspired computing},
  author={Luo, Zhen and Wang, Zijian and Guan, Zeyu and Ma, Chao and Zhao, Letian and Liu, Chuanchuan and Sun, Haoyang and Wang, He and Lin, Yue and Jin, Xi and Yin, Yuewei
and Li, Xiaoguang
},
  journal={Nat. Commun.},
  volume={13},
  number={1},
  pages={699},
  year={2022},
  publisher={Nature Publishing Group UK London}
}

@article{zeng2021electric,
  title={Electric field gradient-controlled domain switching for size effect-resistant multilevel operations in HfO2-based ferroelectric field-effect transistor},
  author={Zeng, Binjian and Liu, Chen and Dai, Siwei and Zhou, Pingan and Bao, Keyu and Zheng, Shuaizhi and Peng, Qiangxiang and Xiang, Jinjuan and Gao, Jianfeng and Zhao, Jie and Liao, Min
and Zhou, Yichun},
  journal={Adv. Funct. Mater.},
  volume={31},
  number={17},
  pages={2011077},
  year={2021},
  publisher={Wiley Online Library}
}

@article{soliman2023first,
  title={First demonstration of in-memory computing crossbar using multi-level cell FeFET},
  author={Soliman, Taha and Chatterjee, Swetaki and Laleni, Nellie and M{\"u}ller, Franz and Kirchner, Tobias and Wehn, Norbert and K{\"a}mpfe, Thomas and Chauhan, Yogesh Singh and Amrouch, Hussam},
  journal={Nat. Commun.},
  volume={14},
  number={1},
  pages={6348},
  year={2023},
  publisher={Nature Publishing Group UK London}
}

@article{boyn2017learning,
  title={Learning through ferroelectric domain dynamics in solid-state synapses},
  author={Boyn, Sören
and Grollier, Julie
and Lecerf, Gwendal
and Xu, Bin
and Locatelli, Nicolas
and Fusil, Stéphane
and Girod, Stéphanie
and Carrétéro, Cécile
and Garcia, Karin
and Xavier, Stéphane
and Tomas, Jean
and Bellaiche, Laurent
and Bibes, Manuel
and Barthélémy, Agnès
and Saïghi, Sylvain
and Garcia, Vincent},
  journal={Nat. Commun.},
  volume={8},
  number={1},
  pages={14736},
  year={2017},
  publisher={Nature Publishing Group UK London}
}

@article{kim2020synergistic,
  title={Synergistic improvement of long-term plasticity in photonic synapses using ferroelectric polarization in hafnia-based oxide-semiconductor transistors},
  author={Kim, Min-Kyu and Lee, Jang-Sik},
  journal={Adv. Mater.},
  volume={32},
  number={12},
  pages={1907826},
  year={2020},
  publisher={Wiley Online Library}
}

@article{ma2020sub,
  title={Sub-nanosecond memristor based on ferroelectric tunnel junction},
  author={Ma, Chao and Luo, Zhen and Huang, Weichuan and Zhao, Letian and Chen, Qiaoling and Lin, Yue and Liu, Xiang and Chen, Zhiwei and Liu, Chuanchuan and Sun, Haoyang and Jin, Xi
and Yin, Yuewei
and Li, Xiaoguang
},
  journal={Nat. Commun.},
  volume={11},
  number={1},
  pages={1439},
  year={2020},
  publisher={Nature Publishing Group UK London}
}

@article{kumar2022high,
  title={High-Performing Self-Powered Photosensing and Reconfigurable Pyro-photoelectric Memory with Ferroelectric Hafnium Oxide},
  author={Kumar, Mohit and Seo, Hyungtak},
  journal={Adv. Mater.},
  volume={34},
  number={5},
  pages={2106881},
  year={2022},
  publisher={Wiley Online Library}
}

@article{geler2022ferroelectric,
  title={A ferroelectric multilevel non-volatile photonic phase shifter},
  author={Geler-Kremer, Jacqueline and Eltes, Felix and Stark, Pascal and Stark, David and Caimi, Daniele and Siegwart, Heinz and Jan Offrein, Bert and Fompeyrine, Jean and Abel, Stefan},
  journal={Nat. Photon.},
  volume={16},
  number={7},
  pages={491--497},
  year={2022},
  publisher={Nature Publishing Group UK London}
}

@article{wang2023ultrathin,
  title={Ultrathin Nitride Ferroic Memory with Large ON/OFF Ratios for Analog In-Memory Computing},
  author={Wang, Ding and Wang, Ping and Mondal, Shubham and Hu, Mingtao and Wu, Yuanpeng and Ma, Tao and Mi, Zetian},
  journal={Adv. Mater.},
  volume={35},
  number={20},
  pages={2210628},
  year={2023},
  publisher={Wiley Online Library}
}

@article{li2024realization,
  title={Realization of sextuple polarization states and interstate switching in antiferroelectric $\mathrm{CuInP_{2}S_{6}}$},
  author={Li, Tao and Wu, Yongyi and Yu, Guoliang and Li, Shengxian and Ren, Yifeng and Liu, Yadong and Liu, Jiarui and Feng, Hao and Deng, Yu and Chen, Mingxing and Zhang, Zhenyu
and Min, Tai
},
  journal={Nat. Commun.},
  volume={15},
  number={1},
  pages={2653},
  year={2024},
  publisher={Nature Publishing Group UK London}
}

@article{jandhyala2013graphene,
  title={Graphene-ferroelectric hybrid devices for multi-valued memory system},
  author={Jandhyala, S and Mordi, G and Mao, D and Ha, M-W and Quevedo-Lopez, MA and Gnade, BE and Kim, J},
  journal={Appl. Phys. Lett.},
  volume={103},
  number={2},
  year={2013},
  publisher={AIP Publishing}
}

@article{li2023gate,
  title={A gate programmable van der Waals metal-ferroelectric-semiconductor vertical heterojunction memory},
  author={Li, Wanying and Guo, Yimeng and Luo, Zhaoping and Wu, Shuhao and Han, Bo and Hu, Weijin and You, Lu and Watanabe, Kenji and Taniguchi, Takashi and Alava, Thomas and Chen, Jiezhi
and Gao, Peng
and Li, Xiuyan
and Wei, Zhongming
and Wang, Lin-Wang
and Liu, Yue-Yang
and Zhao, Chengxin
and Zhan, Xuepeng
and Han, Zheng Vitto
and Wang, Hanwen},
  journal={Adv. Mater.},
  volume={35},
  number={5},
  pages={2208266},
  year={2023},
  publisher={Wiley Online Library}
}

@article{wang2021van,
  title={Van der Waals engineering of ferroelectric heterostructures for long-retention memory},
  author={Wang, Xiaowei and Zhu, Chao and Deng, Ya and Duan, Ruihuan and Chen, Jieqiong and Zeng, Qingsheng and Zhou, Jiadong and Fu, Qundong and You, Lu and Liu, Song and Edgar, James H.
and Yu, Peng
and Liu, Zheng
},
  journal={Nat. Commun.},
  volume={12},
  number={1},
  pages={1109},
  year={2021},
  publisher={Nature Publishing Group UK London}
}

@article{wang2023moire,
  title={Moir{\'e} synaptic transistor for homogeneous-architecture reservoir computing},
  author={Wang, Pengfei and Chen, Moyu and Xie, Yongqin and Pan, Chen and Watanabe, Kenji and Taniguchi, Takashi and Cheng, Bin and Liang, Shi-Jun and Miao, Feng},
  journal={Chin. Phys. Lett.},
  volume={40},
  number={11},
  pages={117201},
  year={2023},
  publisher={Chinese Physical Society and IOP Publishing Ltd}
}

@article{yan2023moire,
  title={Moir{\'e} synaptic transistor with room-temperature neuromorphic functionality},
  author={Yan, Xiaodong and Zheng, Zhiren and Sangwan, Vinod K and Qian, Justin H and Wang, Xueqiao and Liu, Stephanie E and Watanabe, Kenji and Taniguchi, Takashi and Xu, Su-Yang and Jarillo-Herrero, Pablo and Ma, Qiong
and Hersam, Mark C.
},
  journal={Nature},
  volume={624},
  number={7992},
  pages={551--556},
  year={2023},
  publisher={Nature Publishing Group UK London}
}

@article{bian2024developing,
  title={Developing fatigue-resistant ferroelectrics using interlayer sliding switching},
  author={Bian, Renji and He, Ri and Pan, Er and Li, Zefen and Cao, Guiming and Meng, Peng and Chen, Jiangang and Liu, Qing and Zhong, Zhicheng and Li, Wenwu and Liu, Fucai},
  journal={Science},
  volume={385},
  number={6704},
  pages={57--62},
  year={2024},
  publisher={American Association for the Advancement of Science}
}

@article{xu2021non,
  title={Non-volatile multi-level cell storage via sequential phase transition in $\mathrm{Sb_{7}Te_{3}}$/$\mathrm{GeSb_{6}Te}$
  multilayer thin film},
  author={Xu, Zhehao and Su, Xiao and Hua, Sicong and Zhai, Jiwei and Song, Sannian and Song, Zhitang},
  journal={Nanotechnology},
  volume={33},
  number={7},
  pages={075701},
  year={2021},
  publisher={IOP Publishing}
}

@article{wang2025composition,
  title={Composition-Graded Nitride Ferroelectrics Based Multi-Level Non-Volatile Memory for Neuromorphic Computing},
  author={Wang, Rui and Ye, Haotian and Xu, Xifan and Wang, Jinlin and Feng, Ran and Wang, Tao and Sheng, Bowen and Liu, Fang and Shen, Bo and Wang, Ping and Wang, Xinqiang},
  journal={Adv. Mater.},
  volume={37},
  number={5},
  pages={2414805},
  year={2025},
  publisher={Wiley Online Library}
}

@article{cai20258,
  title={8-bit states in 2$\mathrm{D}$ floating-gate memories using gate-injection mode for large-scale convolutional neural networks},
  author={Cai, Yuchen and Yang, Jia and Hou, Yutang and Wang, Feng and Yin, Lei and Li, Shuhui and Wang, Yanrong and Yan, Tao and Yan, Shan and Zhan, Xueying and He, Jun and Wang, Zhenxing},
  journal={Nat. Commun.},
  volume={16},
  number={1},
  pages={2649},
  year={2025},
  publisher={Nature Publishing Group UK London}
}

@article{you2014organic,
  title={An organic multilevel non-volatile memory device based on multiple independent switching modes},
  author={You, Yintao and Yang, Kunlong and Yuan, Sijian and Dong, Shiqi and Zhang, Huotian and Huang, Qinglan and Gillin, William P and Zhan, Yiqiang and Zheng, Lirong},
  journal={Org. Electron.},
  volume={15},
  number={9},
  pages={1983--1989},
  year={2014},
  publisher={Elsevier}
}

@article{rios2015integrated,
  title={Integrated all-photonic non-volatile multi-level memory},
  author={R{\'\i}os, Carlos and Stegmaier, Matthias and Hosseini, Peiman and Wang, Di and Scherer, Torsten and Wright, C David and Bhaskaran, Harish and Pernice, Wolfram HP},
  journal={Nat. Photon.},
  volume={9},
  number={11},
  pages={725--732},
  year={2015},
  publisher={Nature Publishing Group UK London}
}

@article{mallik2023thermally,
  title={Thermally driven multilevel non-volatile memory with monolayer $\mathrm{MoS_{2}}$ for brain-inspired artificial learning},
  author={Mallik, Sameer Kumar and Padhan, Roshan and Sahu, Mousam Charan and Roy, Suman and Pradhan, Gopal K and Sahoo, Prasana Kumar and Dash, Saroj Prasad and Sahoo, Satyaprakash},
  journal={ACS Appl. Mater. Interfaces},
  volume={15},
  number={30},
  pages={36527--36538},
  year={2023},
  publisher={ACS Publications}
}

@article{hyun2016multi,
  title={Multi-floor cascading ferroelectric nanostructures: multiple data writing-based multi-level non-volatile memory devices},
  author={Hyun, Seung and Kwon, Owoong and Lee, Bom-yi and Seol, Daehee and Park, Beomjin and Lee, Jae Yong and Lee, Ju Hyun and Kim, Yunseok and Kim, Jin Kon},
  journal={Nanoscale},
  volume={8},
  number={3},
  pages={1691--1697},
  year={2016},
  publisher={Royal Society of Chemistry}
}

@article{chen2022reproducible,
  title={Reproducible Non-Volatile Multi-State Storage and Emulation of Synaptic Plasticity Based on a Copper-Nanoparticle-Embedded $\mathrm{HfO_{x}}$/$\mathrm{ZnO}$ Bilayer with Ultralow-Switching Current and Ideal Data Retention},
  author={Chen, Shuai and Chen, Hao and Lai, Yunfeng},
  journal={Nanomaterials},
  volume={12},
  number={21},
  pages={3769},
  year={2022},
  publisher={MDPI}
}

\vspace{0.5em}
\noindent\textbf{Acknowledgments}

\noindent The authors are grateful to Jinke Bao and Zhu'an Xu for helpful discussions.This research was supported by the National Key Research and Development Program of China (Grant Nos.2024YFA1408101), National Natural Science Foundation of China (Grant Nos. 12474131, 12304168 and 12404047). This research is supported by ``Pioneer'' and ``Leading Goose'' R$\&$D Program of Zhejiang (Grant No. 2024SDXHDX0007) and Zhejiang Provincial Natural Science Foundation of China (Grant No. LQ23A040009). 
The authors thank the support provided by Yangjian Lin, Pei Sheng, Qike Jiang and Chao Zhang from Instrumentation and Service Center for Physical Sciences (ISCPS) at Westlake University.

\vspace{0.5em}
\noindent\textbf{Author contributions}

\noindent Z.X. and X.Y. contributed equally to this work. X.Y. synthesized the crystal with the help from Z.L.. Z.X. fabricated devices and performed transport 
measurements with the help from X.Y.,T.L. and J.W.  Z.X.,X.Y. and X.L. prepared the figures and wrote the paper. X.Y.and X.L. led the project. All authors contributed to the discussion.

\vspace{0.5em}
\noindent \textbf{Competing interests}

\noindent The authors declare no competing interests.

\vspace{0.5em}
\noindent \textbf{Additional information}

\noindent Supplementary information is available for this paper at the on-line version.



\end{document}